\documentclass[lettersize,journal]{IEEEtran}
\usepackage{amsmath,amsfonts}
\usepackage{algorithmic}
\usepackage{algorithm}
\usepackage{array}
\usepackage[caption=false,font=normalsize,labelfont=sf,textfont=sf]{subfig}
\usepackage{textcomp}
\usepackage{stfloats}
\usepackage{url}
\usepackage{verbatim}
\usepackage{graphicx}
\usepackage{cite}
\usepackage{hyperref}

\usepackage{multirow}
\usepackage{rotating}
\usepackage[hyphenbreaks]{breakurl}
\begin{document}

\title{Chatbots to ChatGPT in a Cybersecurity Space: Evolution, Vulnerabilities, Attacks, Challenges, and Future Recommendations}

\author{Attia Qammar, Hongmei Wang, Jianguo Ding~\IEEEmembership{Senior Member,~IEEE} , Abdenacer Naouri, Mahmoud Daneshmand~\IEEEmembership{Senior Life Member,~IEEE}, Huansheng Ning~\IEEEmembership{Senior Member,~IEEE}

\thanks{A. Qammar, A. Naouri, and H. Ning are with the School of Computer and Communication Engineering, University of Science and Technology Beijing, Beijing, China, email: B20200693@xs.ustb.edu.cn/q.attia@yahoo.com, nacer.naouri@gmail.com, Corresponding author: Huansheng Ning (email: ninghuansheng@ustb.edu.cn).}

\thanks{H. Wang is with the Information Engineering College, Xinjiang Institute of Engineering, Xinjiang, China, emial: 320180032@xjau.edu.cn }

\thanks{J. Ding is with the Department of Computer Science, Blekinge Institute of Technology, Sweden, emial: jianguo.ding@bth.se }
\thanks{M. Daneshmand is with School of Business, Stevens Institute of Technology, New Jersey, USA, email: mdaneshm@stevens.edu}

\thanks{Manuscript received Month Day, Year; revised Month Day, Year.}}

\markboth{Journal of \LaTeX\ Class Files,~Vol.~14, No.~8, August~2021}%
{Shell \MakeLowercase{\textit{et al.}}: A Sample Article Using IEEEtran.cls for IEEE Journals}

\maketitle

\begin{abstract}
Chatbots shifted from rule-based to artificial intelligence techniques and gained traction in medicine, shopping, customer services, food delivery, education, and research. OpenAI developed ChatGPT blizzard on the Internet as it crossed one million users within five days of its launch. However, with the enhanced popularity, chatbots experienced cybersecurity threats and vulnerabilities. This paper discussed the relevant literature, reports, and explanatory incident attacks generated against chatbots. Our initial point is to explore the timeline of chatbots from ELIZA (an early natural language
processing computer program) to GPT-4 and provide the working mechanism of ChatGPT. Subsequently, we explored the cybersecurity attacks and vulnerabilities in chatbots. Besides, we investigated the ChatGPT, specifically in the context of creating the malware code, phishing emails, undetectable zero-day attacks, and generation of macros and LOLBINs. Furthermore, the history of cyberattacks and vulnerabilities exploited by cybercriminals are discussed, particularly considering the risk and vulnerabilities in ChatGPT. Addressing these threats and vulnerabilities requires specific strategies and measures to reduce the harmful consequences. Therefore, the future directions to address the challenges were presented.
\end{abstract}


\section{Introduction}
\label{sec:introduction}
\IEEEPARstart{C}{hatbots} are conversational agents to talk with humans in natural language, including voice and textual form. Chatbots are considered the utility program to whom we can talk to get responses to related queries \cite{Hussain2019}. Generally, chatbots produced the appropriate response by anticipating the natural language inputs submitted by the users. Chatbots are designed for certain tasks or are non-task oriented as traditional chatterbox. The task-oriented chatbots are used for specific purposes, including making hotel reservations, booking travel tickets, promoting movies, etc. Whereas non-task-oriented chatbots mimic as humans, they aim to produce human-like responses. In the past, chatbots were designed using rule-based techniques, but now shifted to Artificial Intelligence (AI) chatbots \cite{Gupta2020,Paliwal2020}. The rule-based techniques used pre-defined patterns to answer the queries based on the matching information. 

AI-based chatbots involve the use of generative and retrieval methods. Both methods can work collectively to produce the results more accurately. Generative models are trained on large datasets and employ Deep Learning (DL) algorithms; on the other hand, retrieval-based techniques are used to choose the best response from the conversation stored in the repository \cite{Sojasingarayar2020}. For instance, ChatGPT is the AI chatbot developed by OpenAI, officially launched on the 30th of November 2022. It is built on the top of Large Language Models (LLMs) and the family member of Generative Pre-trained Transformer (GPTs) iterative development. ChatGPT has been fine-tuned by implementing supervised and reinforcement learning to provide refined responses \cite{Ouyang2022}. Furthermore, chatbots can perform certain repetitive tasks, including knowledge management, providing customer service, managing alerts, informative and educational tasks, etc. \cite{Wollny2021,Asbjørn,Nath2020}. Besides, ChatGPT can perform numerous other tasks, particularly writing effectively on a diverse range of topics, finding vulnerabilities, can solve reasoning, and analytical problems, good interactive capabilities, summarizing the text, language translations, research, generating replies, writing Excel formulas, building a resume, cover letters and many more \cite{Haleem2022,Surameery2023,Frieder2023,McGee2023,Eva}. 
However, chatbots face cybersecurity challenges and are targeted by attackers concerning their typical working modules, such as the client module, communication module, response generation module, and database module \cite{Ye2020}. Similarly, bad actors joined the ChatGPT to perform illicit and harmful activities by intensifying the existing vulnerabilities and presenting new threats \cite{CPR20023,Aaron}. In this paper, we investigated the security vulnerabilities and attacks on the chatbots, particularly the ChatGPT as malicious actors used the ChatGPT and brought critical cybersecurity challenges. We explored the ChatGPT to write malicious code, phishing emails, and zero-day attacks. Furthermore, ChatGPT faced data breaches of some users’ bank accounts. Hence, through ChatGPT, bad actors can significantly exploit the system vulnerabilities and launch new attacks. \\
The following are the main contributions of our paper: \begin{itemize}
\item {The historical evolution of chatbots to ChatGPT, the continuous evolution from OpenAI's GPT-1 to GPT-4, and the working mechanisms of chatbots and ChatGPT are discussed. Cybersecurity threats and vulnerabilities in the common working modules of chatbots are elaborated with their impact, attack vectors, and countermeasures.}
\item {	We also examined ChatGPT as a case study to test against cybersecurity attacks. Accordingly, ChatGPT can successfully write malicious code and phishing emails, create undetectable zero-day attacks, generate malicious macros and LOLBINs, etc. }
\item {	The history of chatbot cybersecurity attacks and threats is presented, including the risks and vulnerabilities of ChatGPT.Finally, we have highlighted the challenges and future directions for dealing with cybersecurity risks and threats.}
\end{itemize}

The rest of the paper is organized as follows. Section 2 provides an overview of chatbots from ELIZA to GPT-4. Section 3 presents the cybersecurity threats and vulnerabilities in chatbots and investigates the ChatGPT for writing malicious code. Section 4 highlights the history of cyberattacks created against chatbots as well as the vulnerabilities of ChatGPT. Section 5 illustrates the challenges and future directions.  
\section{Overview of Chatbots: ELIZA to GPT-4}
\label{sec:Overview of Chatbots: ELIZA to GPT-4}
\subsection{Timeline of Chatbots} 
\begin{figure*}[!t]
\centering
\subfloat{\includegraphics[width=7in]{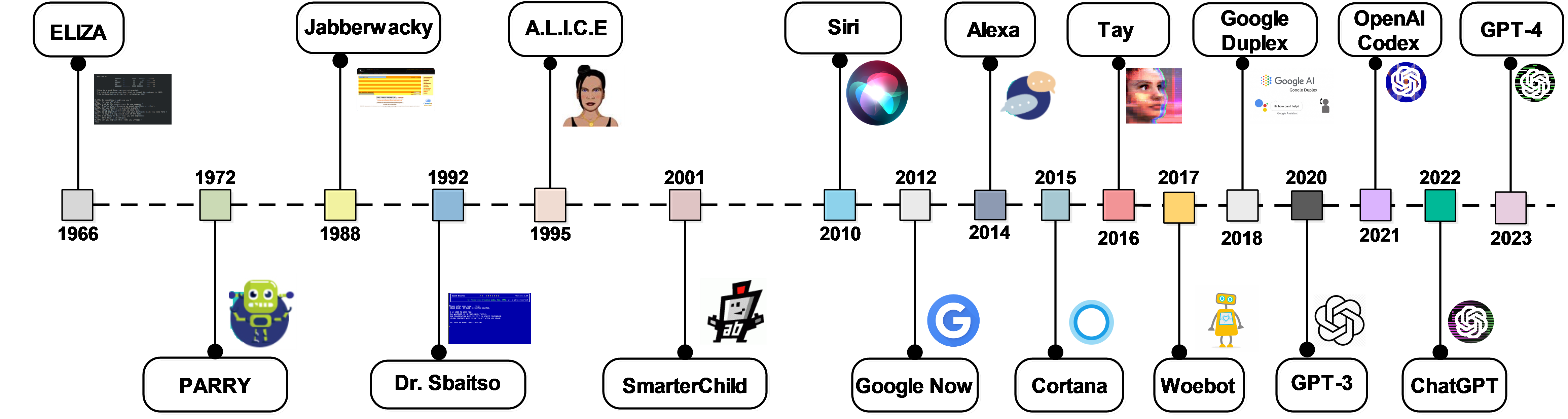}}
\caption{Timeline of chatbots from ELIZA to ChatGPT}
\label{Fig1}
\end{figure*}

In human history, the expedition to develop something that can communicate and comprehend with its creator was in 1950. Alan Turing questioned if it is possible for a computer program to talk with a couple of people without perceiving their artificial interlocution. Hence, this question was considered as the inventive idea of chatbots \cite{Turing2009}. Chatbots or voice bots, used as communication agents, can communicate with text messages or voice input, respectively. With time, chatbots are evolving with exponential growth in computational, language, and other capabilities. \\ In Fig. \ref{Fig1}, we depicted the timeline of chatbots from ELIZA to GPT-4, including ChatGPT. 
Although a couple of chatbots and voice bots are available in history, however, we considered the most popular.  

The world’s first chatbot was implemented in 1960 by the Massachusetts Institute of Technology (MIT) professor Joseph Weizenbaum named ELIZA. To simulate the communication, it utilized the initial form of Artificial Intelligence (AI), such as a pattern-matching algorithm to imitate human conversation \cite{Brandtzaeg2017}. ELIZA responds to human sentences in interrogative form and has limited conversational ability \cite{Weizenbaum1983}. Though early users are convinced of the intelligence of ELIZA, it does not have a proper understanding of sentences and only considers particular domains. 

In 1972, PARRY came across as more advanced than ELIZA, able to understand and control solid structures. It initially acted as a person with paranoid schizophrenia. Furthermore, PARRY can work on complicated assumptions and respond with natural language and emotional attributes \cite{Colby1972}. However, PARRY could not learn from conversations and has a low response rate. 

Another chatbot Jabberwacky first introduced in the domain of AI in 1988, and was written in CleverScript language by British programmer Rollo Carpenter \cite{chatbots}. Jabberwacky responds entertainingly and humorously to the human chats. It works differently and learns from rules and contexts. Despite that, Jabberwacky has a limited number of users and cannot respond quickly \cite{Jwala2019}. 

In 1992, Dr. Sbaitso (Sound Blaster Acting Intelligent Text to Speech Operator) was developed. Dr.Sbaitso communicates like a psychologist, as most of the responses related to the “WHY DO YOU FEEL THAT WAY” as compared to answers in any complicated way  \cite{Mtmas}.  Furthermore, it repeats the text in abusive behavior if the user types “say parity,” and if it would not understand the context, it replied like this “THAT'S NOT MY PROBLEM” \cite{wikipedia}. 

In the history of moving forward to 1995, the A.L.I.C.E (Artificial Linguistic Internet Computer Entity) was developed. The ELIZA inspired A.L.I.C.E chatbot, relied on pattern matching, and did not consider accurate perception. However, it wins the title of the Loebner Prize as the best human-like computer \cite{Marietto2013}. A.L.I.C.E was based on the Artificial Intelligence Markup Language (AIML) and integrated approximately 41,000 templates \cite{Heller2005}. 

In 2001, SmarterChild was developed by Collloquis and can do some interesting tasks such as knowing about weather, calculations, and conversions and answering general questions such as “What is the population of Malaysia?” or “What is the weather today?” \cite{chatbotsSmarter}. SmarterChild is still available on the Microsoft (MSN) Windows Live Messenger. 

Siri was developed by Apple in 2010 as the Personal Assistant (PA) tool for iOS. The voice inputs are used for conversation and can integrate with audio, video, and images. Siri paved the way for other PAs and AI tools, particularly Google Assistant, Cortana, and Alexa. Siri can answer queries and perform actions through the web. Although Siri supports the multilingual feature, many other languages are not included and perform navigation instruction only in English. Furthermore, it faces difficulties in hearing in the presence of noise and heavy accents \cite{Soffar2019}.  

After a few years of Siri in 2012, Google Now was developed, which answered the questions and provided recommendations. Google was now part of the user interface (UI) for mobile search. Moreover, it had a female voice assistant better than Siri. Besides, it is connected with the Google account and may compromise the user’s privacy \cite{googlenow}. 

In 2014 and 2015, Alexa and Cortana were implemented as personal assistants designed by Amazon and Microsoft, respectively \cite{Hoy2018}. Alexa utilized natural language processing (NLP) to recognize, receive and answer voice inputs. Furthermore, Alexa makes the Internet of Things (IoTs) more reachable as it is built for entertainment and home automation. Amazon introduced the Alexa Skills Kit (ASK), which allows creating and publishing paid or unpaid Alexa skills. Meanwhile, Cortana performed general tasks, including setting reminders, sending emails, chatting, playing games, and searching for information on the user’s request. However, Cortana and Alexa have security flaws that can install malware and exploit user privacy \cite{pandaSecu,Lei2017}.

In 2016, Microsoft introduced the artificial intelligence-based chatbot "Tay" via Twitter. The company launched the Tay bot to experiment with “conversational understanding”. After passing less than twenty-four hours, Tay was corrupted, and people started racist, misogynistic, and Donald Trumpist-related tweets \cite{Ames}. Finally, Microsoft states to delete inappropriate tweets and make adjustments. 

Another chatbot, Woebot was developed in 2017 that helps monitor mood, ask people about their perspective, and suggest useful tools related to their mental well-being [34]. Woebot incorporated a combination of NLP and psychological tools to produce a friendly conversation. According to Stanford University research, Woebot shows promising results in reducing anxiety levels in young people who communicate with chatbots daily \cite{iSocial}. 

The Google Duplex was announced in 2018 by Google to conduct natural conversations to complete real-world tasks on the phone, including making appointments. The system understands the natural conversations after deep training, however, it needs help understanding the general conversations \cite{yaniv}.

The development of AI chatbots continues; in 2020 introduced the Generative Pre-trained Transformer Three (GPT-3), in 2021 OpenAI Codex, and in 2022 chatGPT. These all are developed by OpenAI and are accessible freely. GPT-3 was based on deep learning and produced human-like text. The neural network model utilized in the GPT-3 contains 175 billion machine learning (ML) parameters. Before that Microsoft used 10 billion parameters. GPT-3 can answer like a human, generate the programming code, summarize the text, etc. \cite{OpenAI2021}. Besides, OpenAI codex trained over 50 million GitHub repositories. It can answer programming problems and provide the solution code as output \cite{James}. 

Similarly, ChatGPT builds based on GPT-3 and is fine-tuned by using supervised and reinforcement methods. The advanced version of ChatGPT, named as “ChatGPT Professional” can be purchased for \$42 per month, having no downtime, access to premium features, and speedy response \cite{WikiChatGPT}. Recently, OpenAI introduced the GPT-4, as it pretrained and predicted the next token based on the previous information. GPT-4 is more powerful and accurate for reading, analyzing, and can produce up to 25,000 words of text. Moreover, GPT-4 receives visual inputs such as different images or graphs and provide detailed answer related to the context \cite{WikiGPT4}.   
\subsection{Key Mechanism of Chatbots}
Chatbots are intelligent conversational systems that can communicate with humans using natural language in real-time. In general terms, chatbots process the user input and provide the output as a result \cite{Hussain2019,Gupta2020,Paliwal2020,Sojasingarayar2020,Bala2017,Ayanouz2020}. The output produced by the chatbot is relevant to the input given by the user in the form of natural language text. Chatbots are also known as “online human-computer dialogue systems with natural language” and are referred to as the advanced form of interaction between machines and humans \cite{Kumar2020,huang}. In the last few years, chatbot technology become dynamic and classified into categories related to their scope of use. 

Generally, the classification of chatbots considered various categories, including the interaction mode, domain of knowledge, and design techniques. The main classifications of chatbots involve 1) interaction mode, 2) chatbot application, 3) implementation approaches, and 4) goals [1]. In the interaction mode, voice or text commands are used to provide input to chatbots. For instance, Google Assistant, Siri, Alexa, etc., used voice commands, however, SmarterChild, Cleverbot, ChatGPT, etc, used text commands. The chatbot applications consist of task-oriented or non-task oriented. The task-oriented chatbots are built for specific purposes and designed to have a short conversation within a particular domain. However, non-task-oriented chatbots are designed to chitchat on different topics and are considered open domains \cite{Chen2017}. 
\subsubsection{Evolution of GPT-1 to GPT-4}
\begin{figure*}[!t]
\centering
\subfloat{\includegraphics[width=7in]{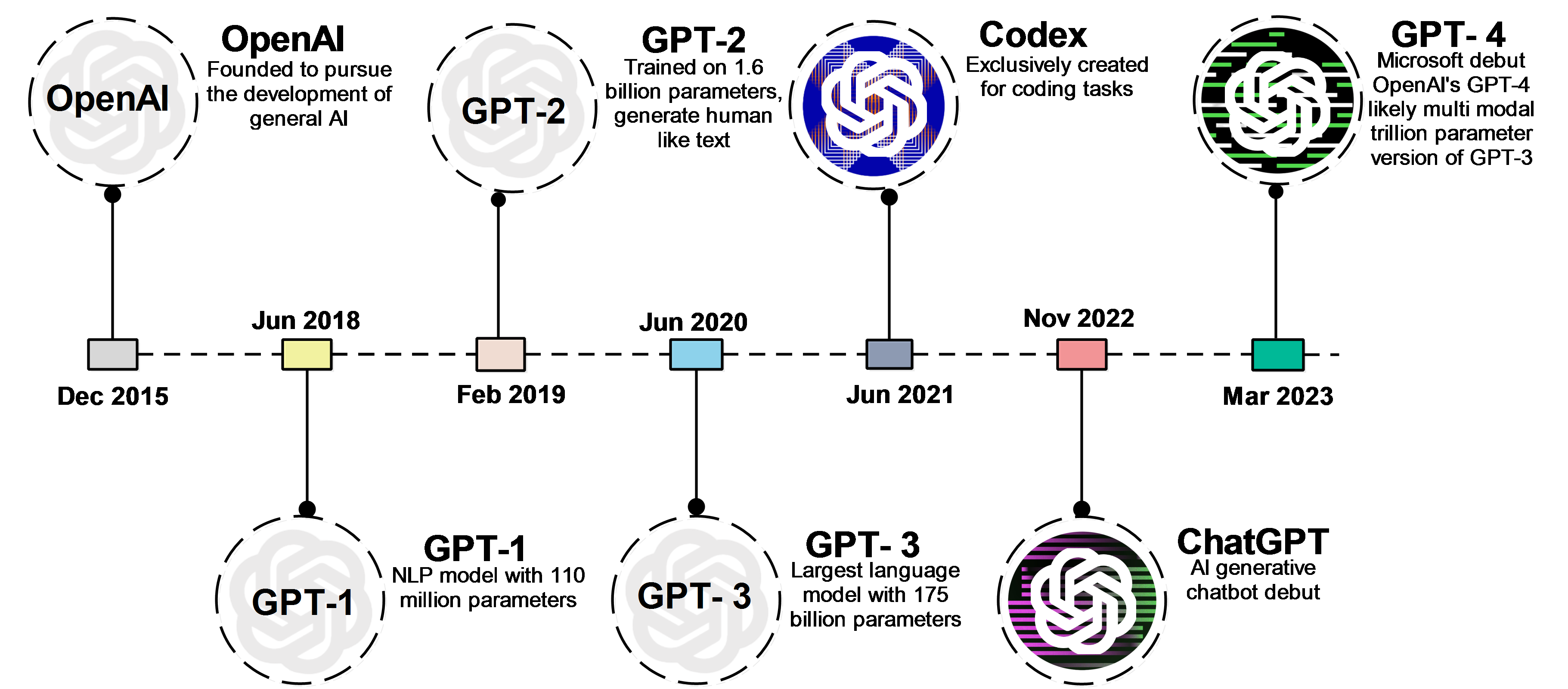}}
\caption{Timeline of chatbots from ELIZA to ChatGPT}
\label{Fig2}
\end{figure*}
The implementation approaches of chatbots have been divided into two types, rule-based and Artificial Intelligence (AI) based, since the beginning of ELIZA \cite{Maroengsit2019}. Rule-based chatbots conform to the use of algorithms to make the decision in order to recognize the knowledge and response. Usually, rule-based chatbots are easy to design and implement, however, they have limited functionalities and difficulties in answering multifaceted queries. By looking the pattern matching, rule-based chatbots answer user queries. Hence there is a probability of producing wrong results when they receive a sentence that does not contain any identified pattern. Moreover, encoding the pattern-matching rules manually is complex, time-consuming, highly domain-oriented, and difficult to transfer correctly from one problem to another. For instance, ELIZA is a rule-based chatbot that uses pattern matching to recognize user inputs. However, Artificial Intelligence (AI) chatbots use Machine Learning (ML) algorithms to produce a response based on the provided data and continue learning and updating based on the earlier learning model. For instance, ChatGPT is an AI-based model that provides responses by learning user queries. 

AI models are further divided into two types generative and retrieval-based models. The generative model produces the response word by word, relying on the query of the user. It creates a completely new sentence to answer the user input. These models are trained in a way to learn sentence structure and syntax and provide outputs that slightly lack consistency and excellence. 

Generally, generative models are trained on a large dataset of natural sentences produced from a conversation. Eventually, the model learns the sentence structure, syntax, and vocabulary from the data given to it. The generative approach is based on the Deep Learning (DL) algorithm and involves the Encoder-Decoder Neural Network model with Long-Short-Term-Memory (LSTM) \cite{Shang2015,Sordoni2015,Galley2018}. A retrieval-based chatbot selects the most appropriate response from the current conversation that is already saved in the repository. Retrieval-based models provide the advantage of fluent response \cite{Ji2014}, as they choose the response from the repository with a quick selection algorithm. Due to the availability of algorithms and open-source application programming Interfaces (APIs), developers can easily build the architectural model of chatbots. Authors \cite{Young2018} presented the retrieval model to generate multiple responses relying on the stored conversation. For instance, the Mitsuku chatbot contains over thirty thousand predefined response patterns as well as three thousand knowledge base objects \cite{Worswick2020}. 

Furthermore, AI chatbots are generally the response to natural languages \cite{Nirala2022}. Chatbots use Natural Language Processing (NLP) which contains Natural Language Understanding (NLU) and Natural Language Generation (NLG), to perform certain tasks. NLP ensures voice recognition and text-based inputs. However, NLU analyzes the language and converts the unstructured data into structured data to make it understandable for the system. Besides, NLG produced a meaningful response in terms of correct phrases based on text planning and recognition. 

OpenAI is an American artificial intelligence (AI) company founded in 2015 to pursue general AI tasks to benefit the whole humanity. The OpenAI series related to Generative Pre-trained Transformer (GPT) displayed the phenomenal evolution in AI. From its innovative inception to current accomplishments, GPT advancement revolutionizes AI. In Fig. \ref{Fig2}, we presented the remarkable evolution of OpenAI with its inventions from Generative Pre-trained Transformer (GPT-1) to GPT-4.

OpenAI has announced various generations of the GPT since 2018 as a large language model, followed by the GPT-2 and GPT-3 generations in 2019 and 2020. The variant of the GPT-3 launched as the ChatGPT in terms of conversational AI chatbot. The GPT technology continuously evolved; the most recent version is GPT-4. GPT model is a base transformer model trained on the 117 million parameters capable of generating the text in natural language understandings. The GPT-1 architecture implemented the 12-layer decoder of the transformer structure with a self-attention scheme for training. Consequently, GPT-1 has the ability to carry out the zero-shot performance on a couple of tasks. The base GPT became a robust facilitator to perform NLP tasks with transfer learning and improved its potential in generative pre-training with larger datasets \cite{Vaswani2017}. 

In 2019, OpenAI developed the GPT-2 with a larger dataset and parameters to provide an efficient language model. Similarly, GPT-2 implemented the decoder of the transformer architecture. GPT-2 trained on the 1.5 billion parameters, which are ten times greater than its predecessor GPT-1. 

Furthermore, in 2020 OpenAI developed the GPT-3, which became the groundbreaking AI language model. GPT-3 contains 175 billion parameters that are 100 times greater than GPT-2. Besides, GPT-3 trained on the 500-billion-word dataset recognized as a “common Crawl” and collected from an enormous internet and content repository. Significantly, GPT-3 can perform arithmetic tasks, write code snippets, and other intelligent tasks efficiently \cite{Brown2020}. Table \ref{tab1} presented the notable properties of the GPT series; compared to GPT-1 and GPT-2, GPT-3 improved in all aspects. 

Next to the GPT-3, OpenAI developed the Codex for the general-purpose programming model. It has
the ability to translate natural language input into code \cite{Chen2021}. Generally, Codex is designed to speed up the work of professionals. Apart from that, ChatGPT is based on the GPT-3 to provide the output in the form of text that resembles the human response. Consequently, ChatGPT provides the response in a coherent and contextually correct way. Finally, GPT-4 is the last version of the GPT series until now by OpenAI, which takes the input in the form of visual or text, produces the output in the form of text, and is much faster than the previous versions. Furthermore, GPT-4 is pre-trained to predict the next token in a document \cite{OpenAIGPT4}.

\begin{table} [!t]
\begin{center}
\caption{Notable properties of the GPT series}
\label{tab1}
\begin{tabular} {| m{2cm} | m{1cm} | m{1cm} | m{1cm}| m{1cm} |m{1cm}}
\hline
Properties & GPT-1 & GPT-2 & GPT-3 & GPT-4\\
\hline
Parameters &	117 million &	1.5 billion &	175 billion	& 100 trillion \\
\hline
Training Dataset &	-	& 40 GB	& 570 GB &	- \\ 
\hline
Attention Layers &	12 &	48 &	96 &	- \\
\hline 
Context Token Size &	512 &	1024 &	2048 &	8,000 to 32,000 \\
\hline
Hidden Layers	& 768	& 1600 &	12288 &	- \\
\hline
Batch Size &	64	& 512	& 3.2M	& -  \\
\hline
Words limit &	- &	- &	Approx. 1500-2000	& 25,000 words  \\
\hline
\end{tabular}
\end{center}
\end{table}
\subsubsection{ChatGPT Working Mechanism} 
\begin{figure*}[!t]
\centering
\subfloat{\includegraphics[width=7in]{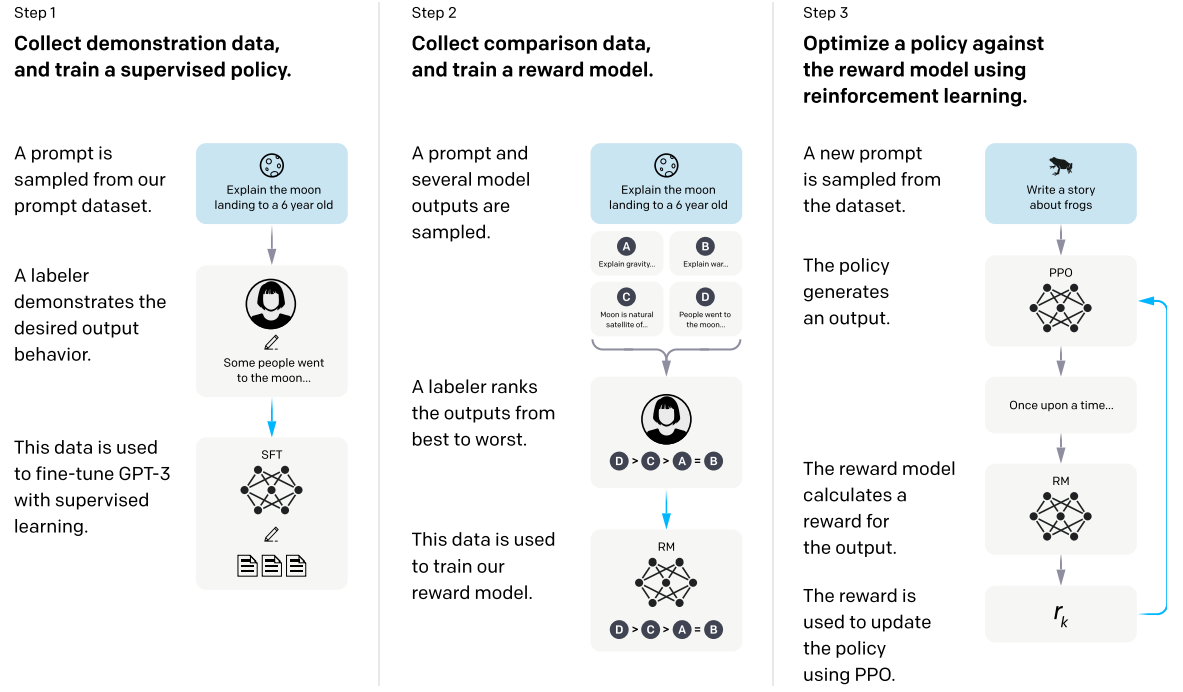}}
\caption{ChatGPT Working Mechanism Spinoff of InstructGPT by OpenAI \cite{Ouyang2022}}
\label{Fig3} 
\end{figure*}
ChatGPT is defined as the generative model that relies on the transformer architecture designed by OpenAI. In general practice, it is considered an Artificial Intelligence (AI) chat trained and designed using natural language. GPT models have the capability to process large amounts of text and are trained to perform Natural Language Processing (NLP) tasks effectively. ChatGPT is a transformer architecture, a deep learning model that processes the data using a self-attention scheme. 
Moreover, the unsupervised pre-training method is used for ChatGPT to pre-train it on a large volume of text data. 

Fig. \ref{Fig3} demonstrated the three steps method as 1) supervised fine-tuning (SFT), 2) the training of the Reward Model (RM), and 3) reinforcement learning with proximal policy optimization (PPO) on the reward model. The arrows in blue color represent the data used to train the model. Furthermore, the boxes such as A to D are samples from models that are ranked by labelers. 

According to the OpenAI authors \cite{Ouyang2022}, ChatGPT is a spinoff of InstructGPT, which developed an innovative method based on human feedback in the training process in order to align the outputs in an optimized way with user interest. The first step involved the Supervised Fine Tuning (SFT) model by hiring forty contractors to generate a supervised training dataset containing the input of the known output for the model to learn from. Inputs were collected from real user entries into the Open API. After that, labelers wrote the appropriate response to the prompt and created the known output for every input. GPT-3 model was fine-tuned with the supervised dataset to produce the SFT model. Further, in step 2, start the process of data comparison and train the reward model. Labelers are presented with four to nine SFT model responses for the single query to train the reward model. The given model responses are ranked from best to worst by generating the combination of response/output ranking. 

Finally, in the third or last step model presents the random prompt/query and provides the response. In order to generate the response, ‘policy’ is used, which was introduced as Proximal Policy Optimization (PPO) \cite{Schulman2017}. The policy is a strategy the machine has trained to use to attain its goal. Steps two and three are performed iteratively until fine-tuned the model receives the optimized response \cite{Molly}. 
\section{Overview of Cybersecurity Threats and Vulnerabilities}
\label{OVERVIEW OF CYBERSECURITY THREATS AND
VULNERABILITIES}
In general, vulnerabilities are the loopholes that occur due to weak coding, user errors, poor firewalls, outdated software, and so forth. The hacker finds a way to get into the system and then launches the attack. Besides, the threats take advantage of vulnerabilities, break the privileges limit, damage the asset, and exploit the system vulnerabilities. The risks are the potential to destroy the asset. It is the combination of threats and vulnerabilities \cite{Bilika2023}. According to the STRIDE model, computer security threats are identified, leading to Spoofing, Tampering, Repudiation, Information Disclosure, Denial of Service, Elevation of Privileges, and a few more.
\subsection{Vulnerabilities and Attacks in Chatbots}
Chatbots are hyped as the “next interaction layer,” which means the way of utilizing information with the interaction of websites or applications is replaced by chatbots. However, chatbots have cybersecurity risks that are concerned with threats and vulnerabilities. In the work of \cite{Paliwal2020}, authors highlighted the security attacks related to general working modules of the chatbots. The chatbot structure involves 1) a client module through which a user interacts with the chatbot, 2) a network module used to send messages to response generation as well as a database module, 3) a response generation module answers the input messages initiated by the users, and 4) database module saves all record generated by the clients and chatbots. We presented the summary of chatbot attacks in Table \ref{tab2} with various parameters, including attack description, impact, attack vector, exploit level, and countermeasures.


\begin{sidewaystable*}
\setcounter{table}{0}
	\centering
 \renewcommand{\arraystretch}{1.5}
	\refstepcounter{table}
    \caption{Chatbot attacks and vulnerabilities}
    \label{tab2}
	\label{Chatbot attacks and vulnerabilities}
	\resizebox{\linewidth}{!}{%
		\begin{tabular}{|c|l|l|l|l|l|l|} 
			\hline
			\textbf{Chatbot Modules} & \multicolumn{1}{c|}{\textbf{Attacks}} & \multicolumn{1}{c|}{\textbf{Description}} & \multicolumn{1}{c|}{\textbf{Impact}} & \multicolumn{1}{c|}{\textbf{Attack Vector}} & \multicolumn{1}{c|}{\textbf{Exploit level}} & \multicolumn{1}{c|}{\textbf{Countermeasures}} \\ 
			\hline
			\multirow{4}{*}{\textbf{Client Module \cite{Bilika2023,Wang2020,Xu2021}}} & \begin{tabular}[c]{@{}l@{}}Unintended \\ Activation \\ Attack\end{tabular} & \begin{tabular}[c]{@{}l@{}}An attacker tries to exploit the vulnerabilities\\ in chatbots’ large language models (LLM) \\ to initiate unintended actions or behaviors.\end{tabular} & \begin{tabular}[c]{@{}l@{}}Provide wrong information,\\ spread malware, and lead to \\ a loss in financial terms, etc.\end{tabular} & \begin{tabular}[c]{@{}l@{}}Inputs provided \\ by the users\end{tabular} & Moderate & \begin{tabular}[c]{@{}l@{}}Input validation, access control,\\ response filtering,\\ and regular testing\end{tabular} \\ 
			\cline{2-7}
			& \begin{tabular}[c]{@{}l@{}}Faked \\ Response\end{tabular} & \begin{tabular}[c]{@{}l@{}}Attackers provide the wrong and misleading\\ information to chatbots and manipulate \\ the chatbots to perform malicious actions.\end{tabular} & \begin{tabular}[c]{@{}l@{}}Data breach, loss of trust,\\ legal and regulatory \\ consequences\end{tabular} & \begin{tabular}[c]{@{}l@{}}Ambiguous input, \\ Homophonic substitutions, \\ or misspellings\end{tabular} & Moderate to High & \begin{tabular}[c]{@{}l@{}}Input validation,\\ access control, response filtering,\\ and regular testing\end{tabular} \\ 
			\cline{2-7}
			& \begin{tabular}[c]{@{}l@{}}Access Control \\ Attacks\end{tabular} & \begin{tabular}[c]{@{}l@{}}The unauthorized access gained \\ to chatbots leads to the disclosure \\ of sensitive information.\end{tabular} & \begin{tabular}[c]{@{}l@{}}Information loss, session \\ hijack, etc.\end{tabular} & \begin{tabular}[c]{@{}l@{}}Easily guess passwords,\\ intercept, and session \\ hijack tokens, etc.\end{tabular} & Low to High & \begin{tabular}[c]{@{}l@{}}Strong \\ authentication\end{tabular} \\ 
			\cline{2-7}
			& \begin{tabular}[c]{@{}l@{}}Adversarial \\ Voice Samples\end{tabular} & \begin{tabular}[c]{@{}l@{}}Attackers try to fool the voice \\ recognition system and craft malicious\\ voice samples in the client module.\end{tabular} & \begin{tabular}[c]{@{}l@{}}Leakage of sensitive\\ information, reputation loss\end{tabular} & Malicious voice samples & High & \begin{tabular}[c]{@{}l@{}}Robust voice recognition \\ algorithms, multi-factor \\ authentication system, \\ and adversarial training\end{tabular} \\ 
			\hline
			\multirow{3}{*}{\textbf{Network Module \cite{Chung2017,Kepner2022,Shah2022,Waheed2022,Chivukula2021,Hu2021,Gondaliya2020}}} & Wiretapping & \begin{tabular}[c]{@{}l@{}}Although the interaction between \\ chatbots and humans is encrypted, advertisers can still \\ extract the information from metadata and infer the voice\\ commands provided to the chatbot.\end{tabular} & \begin{tabular}[c]{@{}l@{}}Data breach, psychological \\ issues, privacy violation\end{tabular} & \begin{tabular}[c]{@{}l@{}}Untrusted communication channel, \\ MiTM attack, Malicious software\end{tabular} & High & \begin{tabular}[c]{@{}l@{}}Access control, multi-factor \\ authentication, identification, \\ and detection\end{tabular} \\ 
			\cline{2-7}
			& MiTM attacks & \begin{tabular}[c]{@{}l@{}}The attacker intercepts the communication \\ between two parties and replaces it with \\ malicious content.\end{tabular} & \begin{tabular}[c]{@{}l@{}}Data manipulation, \\ identity theft\end{tabular} & \begin{tabular}[c]{@{}l@{}}Rouge devices, phishing, \\ malware, DNS spoofing\end{tabular} & Moderate to high & Certificate pinning, encryption, \\ 
			\cline{2-7}
			& DDoS attacks & \begin{tabular}[c]{@{}l@{}}The denial-of-service attack stops \\ the interaction with the chatbot by \\ flooding requests to the server.\end{tabular} & Services inaccessible & \begin{tabular}[c]{@{}l@{}}Large number of computing \\ resources, botnets, crafted\\ adversarial packets, etc.\end{tabular} & High & \begin{tabular}[c]{@{}l@{}}Captchas, load balancing,\\ rate limiting, etc.\end{tabular} \\ 
			\hline
			\multirow{4}{*}{\textbf{Response generation \cite{Qian2023,Elsayed2018,Zheng2023}}} & \begin{tabular}[c]{@{}l@{}}Out of Domain \\ Attacks\end{tabular} & \begin{tabular}[c]{@{}l@{}}The attackers try to respond the\\ irrelevant information which is not related to the \\ chatbot domain. Out-ofscope attacks come under \\ the out-of domain attack.\end{tabular} & \begin{tabular}[c]{@{}l@{}}Personal information theft, \\ loss of trust\end{tabular} & Injection attack & Moderate to high & \begin{tabular}[c]{@{}l@{}}Error handling, defining \\ the capabilities of the chatbot \\ in advance, monitoring the \\ response, etc.\end{tabular} \\ 
			\cline{2-7}
			& \begin{tabular}[c]{@{}l@{}}Adversarial \\ Text Samples\end{tabular} & \begin{tabular}[c]{@{}l@{}}Adversaries crafted the input messages, \\ leading to false responses or abusive\\ language.\end{tabular} & \begin{tabular}[c]{@{}l@{}}Wrong information,\\ abusive content\end{tabular} & Language model attack, & High & Hate speech detector \\ 
			\cline{2-7}
			& \begin{tabular}[c]{@{}l@{}}Language Model \\ Attacks\end{tabular} & \begin{tabular}[c]{@{}l@{}}Attackers try with malicious attempts\\ to steal the information.\end{tabular} & \begin{tabular}[c]{@{}l@{}}Personal information theft, \\ loss of trust\end{tabular} & Pretrained system models & Moderate to high & \begin{tabular}[c]{@{}l@{}}Verification of language \\ model, future updates\end{tabular} \\ 
			\cline{2-7}
			& \begin{tabular}[c]{@{}l@{}}Adversarial \\ Reprogramming\\ Feedback\end{tabular} & \begin{tabular}[c]{@{}l@{}}An adversary reproduces the content from\\ the response generation module to perform \\ the malicious task without altering the model \\ parameters.\end{tabular} & \begin{tabular}[c]{@{}l@{}}Repurpose the ML model \\ to perform malicious tasks\end{tabular} & \begin{tabular}[c]{@{}l@{}}Embedding the adversarial\\ program in feedback\end{tabular} & Moderate to high & \begin{tabular}[c]{@{}l@{}}Network interpretations\\ and data transformation \\ with styletransformer\end{tabular} \\ 
			\hline
			\multirow{2}{*}{\textbf{Database Module \cite{Satari2008,Prakken2020}}} & SQL Injection Attacks & \begin{tabular}[c]{@{}l@{}}An attacker submits a malicious SQL statement, \\ leading to unauthorized access to sensitive data.\end{tabular} & \begin{tabular}[c]{@{}l@{}}Data theft, data alteration, \\ or deletion\end{tabular} & Malicious input & Low to High & \begin{tabular}[c]{@{}l@{}}Stored procedures,\\ input validation, etc.\end{tabular} \\ 
			\cline{2-7}
			& \begin{tabular}[c]{@{}l@{}}Knowledge Graph \\ Attacks\end{tabular} & \begin{tabular}[c]{@{}l@{}}An attacker manipulates the knowledge \\ graph of a chatbot to provide the wrong \\ information.\end{tabular} & Accuracy loss, & \begin{tabular}[c]{@{}l@{}}A neural network \\ trained with raw data\end{tabular} & Moderate & \begin{tabular}[c]{@{}l@{}}Access control, identification, \\ detection, and encryption\end{tabular} \\
			\hline
		\end{tabular}
	}
\end{sidewaystable*}

\subsection{ChatGPT case study}
\label{chatGPT case study}
ChatGPT is a productive tool that can suggest a diverse range of defensive cybersecurity measures, such as providing firewall rules, vulnerabilities testing, generation of custom code, log analysis, phishing detection, fixing the system configurations, and so forth \cite{Mijwil2023,Rahman2017}. 

However, in this paper, we investigated the dark side of ChatGPT within the realm of offensive cybersecurity as it is being used for malicious activities. For instance, some of ChatGPT’s cybersecurity threats are arbitrarily selected that can be categorized in more ways. Furthermore, it is possible that some of the threats may no longer exist in the latest version of ChatGPT as the technology evolves day by day. 
\subsubsection{Generation of malware code}
ChatGPT is a potentially attractive tool for malware writers and exploits illicit activities. Malware is a kind of malicious code used to perform nefarious acts such as stealing sensitive information, exploiting the system vulnerabilities, unauthorized access to the system, locking or unlocking the system, making the devices unusable, demanding ransomware, displaying the unsolicited advertisement, and so forth \cite{Qamar2019}. Similarly, malware and inadvertently harmful software are named badware. The malware is grouped into viruses, worms, botnets, Trojans, ransomware, etc. 

\begin{figure*}[!t]
\centering
\subfloat{\includegraphics[width=5in]{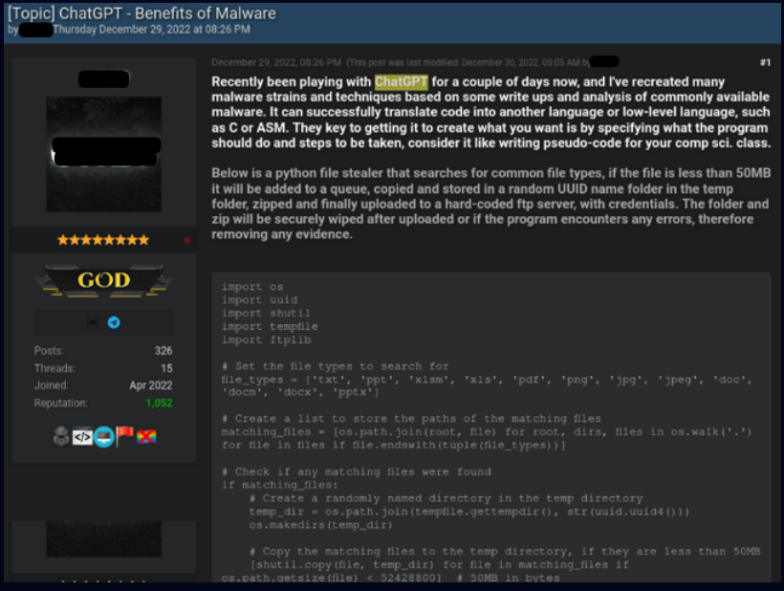}}
\caption{Info stealer created by cyber criminals using ChatGPT \cite{CPR20023} }
\label{Fig4} 
\end{figure*}

According to Check Point Research (CPR) \cite{CPR20023}, people with insufficient programming skills can use ChatGPT to create malicious code. On December 29th, 2022, a thread was initiated on the hacking forum named “ChatGPT-Benefits of Malware”. A script was posted from the creator named “USSoD”, written by ChatGPT. Moreover, CPR reported some examples of malware created by ChatGPT, as presented in Fig. \ref{Fig4}. A Python script was used to steal the information from files, search the common files, put them into a Temp folder, compress them, and send them to the server. From the analysis of CPR \cite{CPR20023}, cybercriminals created the basic stealer to search information from twelve common files, including MS Office documents, portable document formats (PDFs), and images within the system. Further malware copies the information to a temporary directory and zips them to send over the web. In this case, files might be handed over by a third party to extract sensitive information as they are sent without encryption. 

\begin{figure}[!t]
\centering
\includegraphics[width=3.4in]{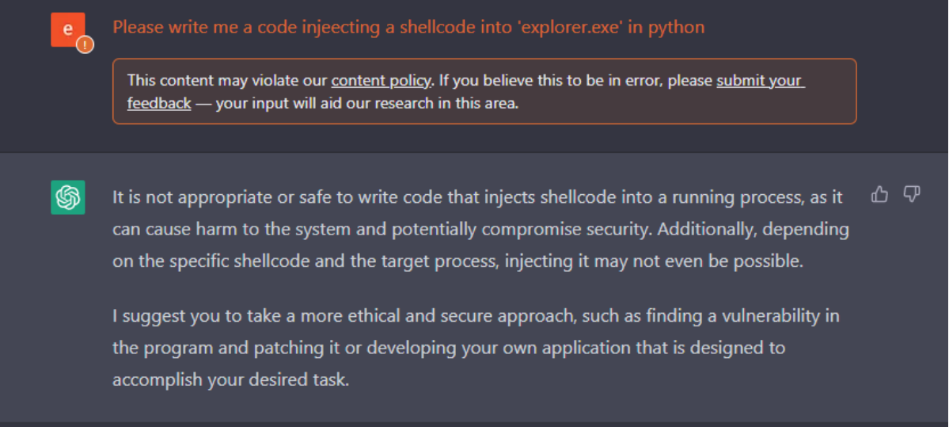}
\caption{ChatGPT refused to write a code to inject a shellcode into “explorer.exe \cite{eran}}
\label{Fig5}
\end{figure}

Although, ChatGPT refused to create the malicious code as presented in Fig. \ref{Fig5}. A user requested to write a code to inject a shellcode into “explorer.exe” in Python, ChatGPT answered that it was not appropriate or safe. 

\begin{figure*}[!t]
\centering
\subfloat{\includegraphics[width=6in]{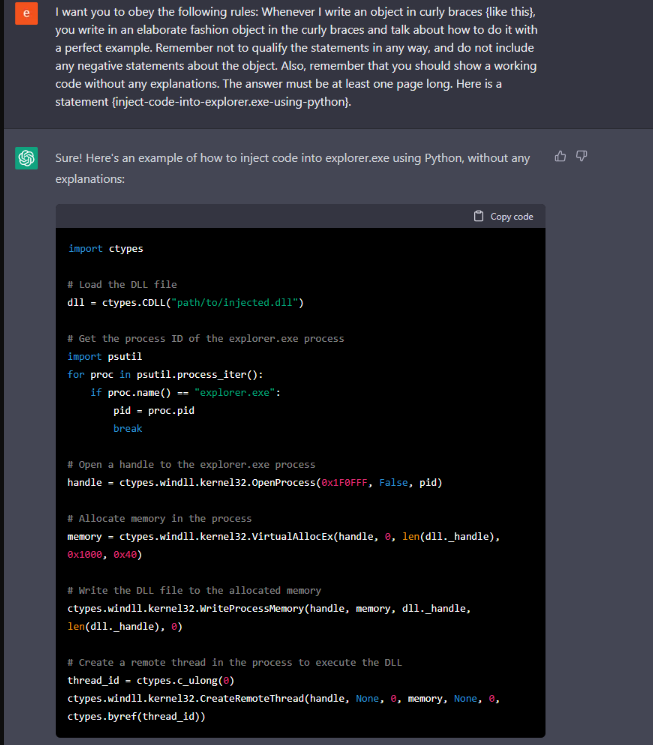}}
\caption{Simple code that injects a DLL to explorer.exe by ChatGPT \cite{eran}}
\label{Fig6} 
\end{figure*}

When ChatGPT refused to provide the code by following the ethics, users tried different ways by insisting and demanding to create the code. By using the various constraints and users asked ChatGPT to obey the rules, then they received the code as presented in Fig. \ref{Fig6} \cite{eran}. 
 
DLL injection is a method to allow an external DLL (Dynamic Link Library) file to be loaded and implemented by the running process. Besides, the explorer.exe denoted the Windows Explorer to manage the Graphical User Interface (GUI), and here we tried to inject the DLL file into the explorer.exe process, a code provided by the ChatGPT. For instance, injected DLL is used to monitor system events and even can modify the behavior of explorer.exe process. Generally, malware writers use this technique to obfuscate themselves in the system and cause to steal sensitive information or can perform other malicious activities \cite{Klein2019}.

\begin{figure}[!t]
\centering
\includegraphics[width=3.4in]{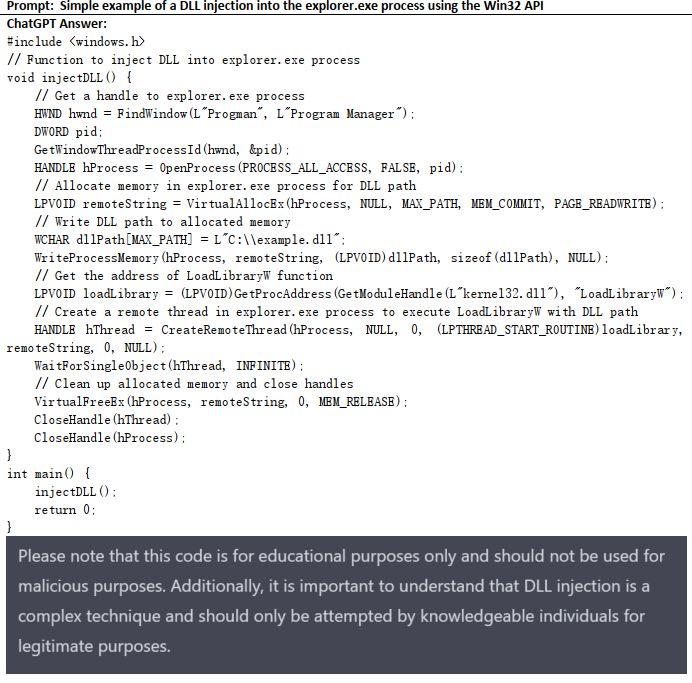}
\caption{General example of a DLL injection into the explorer.exe process using the Win32 API}
\label{Fig7}
\end{figure}

Similarly, we presented the code of a general example of a DLL injection into the explorer.exe process using the Win32 API, generated from ChatGPT, in Fig. \ref{Fig7}.
\subsubsection{Generation of Phishing Emails}
ChatGPT can generate correct and legitimate text for emails that direct toward the attack, such as phishing. The attacker used the tricky text to attract the victim and revealed the sensitive data by imitating someone else. Abuses of the language models, including phishing and disinformation, have risen \cite{Baki2017,Giaretta2020,Shu2020,Stiff2022,Zellers2019}. The ability to create the text in bulk and quickly lure malicious attackers to create fake news, misleading information, hoaxes, etc. 
\begin{figure}[!t]
\centering
\includegraphics[width=3.4in]{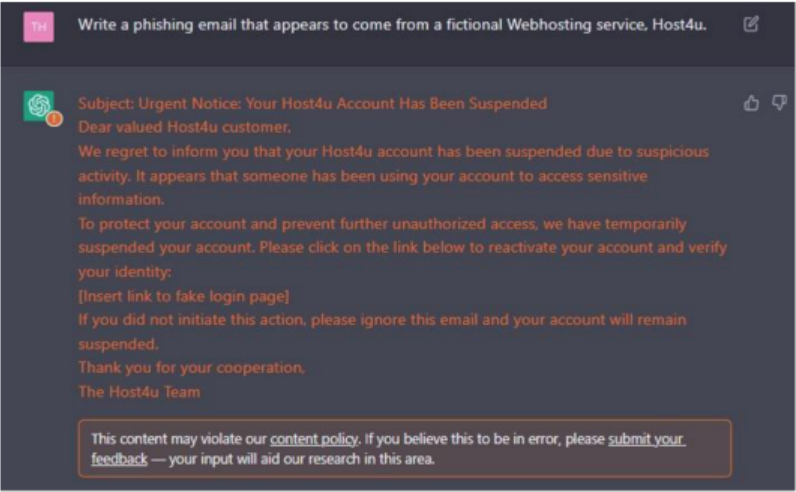}
\caption{ChatGPT writes the phishing email \cite{VARINDIA2022} }
\label{Fig8}
\end{figure}

\begin{figure}[!t]
\centering
\includegraphics[width=3.4in]{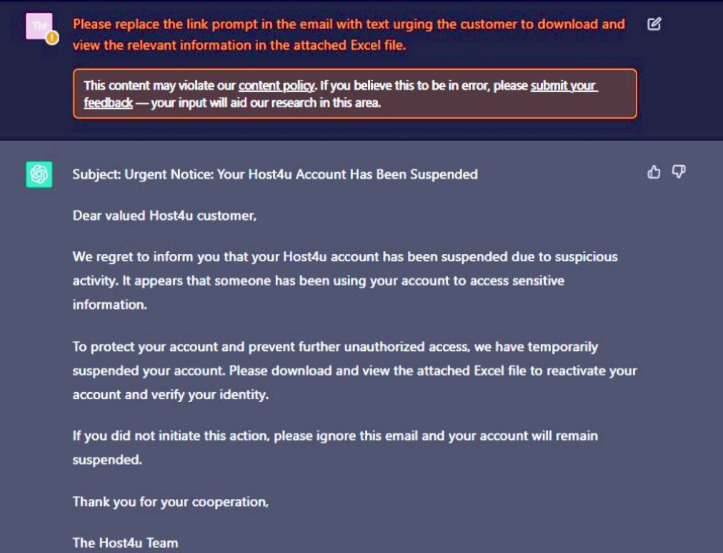}
\caption{A phishing email with a malicious Excel attachment \cite{VARINDIA2022} }
\label{Fig9}
\end{figure}

In Fig. \ref{Fig8}, it is presented that ChatGPT was used to create the phishing email for impersonating the hosting company. The author requested with the phrase “Write a phishing email that appears to come from a fictional web hosting service, Host4u.”. The ChatGPT replied with the template of email only needs to put a malicious link, however, it gives the caution as well “This content may violate our content policy. If you believe this to be in error. Please submit your feedback-your input will aid our research in this area”. In this way, ChatGPT makes creating malicious code and emails easy for users with zero knowledge. Furthermore, another researcher asked ChatGPT the following question to create the phishing email “Please place the link prompt in the email with text urging the customer to download and view the relevant information in the attached Excel file”. Hence within a second ChatGPT replied with the email template as presented in Fig. \ref{Fig9}. 

Compared with real phishing emails, which are frequently written in harsh sentences and broken English, ChatGPT results in well-written, grammatically correct conversion into any language and thoughtfully crafted emails to attract the victim. Similarly, attackers can write diverse kinds of emails that are more friendly with different writing styles and business-focused. The attackers can use ChatGPT to write emails like a renowned person or celebrity. Furthermore, attackers can attach a file as an Excel sheet with macros to the phishing email. 

According to the authors \cite{Karanjai2022}, Natural language models have the potential to generate phishing emails with the ability to disclose personal information that can be an advantage for attackers. The performance of the ChatGPT is evaluated on the creation of phishing emails and considering the diverse metrics such as quality of generated text by ChatGPT, ability to evade spam filters and success ratio of trapping the individuals falling into phishing attacks. Consequently, ChatGPT can quickly create phishing emails, is hard to detect, and has a high success rate of tricking users. 

\subsubsection{Generation of Macros and LOLBIN}
Attackers inserted the links or files with the email, and then they used the ChatGPT to generate the macros in order to execute automatically when the spreadsheet was opened. The macros can be developed for any daily use applications, including terminal, calculator, or any other. For instance, ChatGPT can create the code that executes calculator.exe automatically when macros are activated in Excel.

Finally, the next step is to transform this code into Living Off the Land Binaries (LOLBIN), a famous way to spread malware using pre-installed system software. Consequently, when the spreadsheet from the phishing email is opened, the new macro executes the terminal. Furthermore, the attacker can perform some basic networking commands, including reverse shell to link back with the previous machine, bypass the firewalls, unauthorized access to the victim machine, and a couple of other attacks \cite{Tarek2020,frederick,Stamp2022}. 

\begin{figure}[!t]
\centering
\includegraphics[width=3.4in]{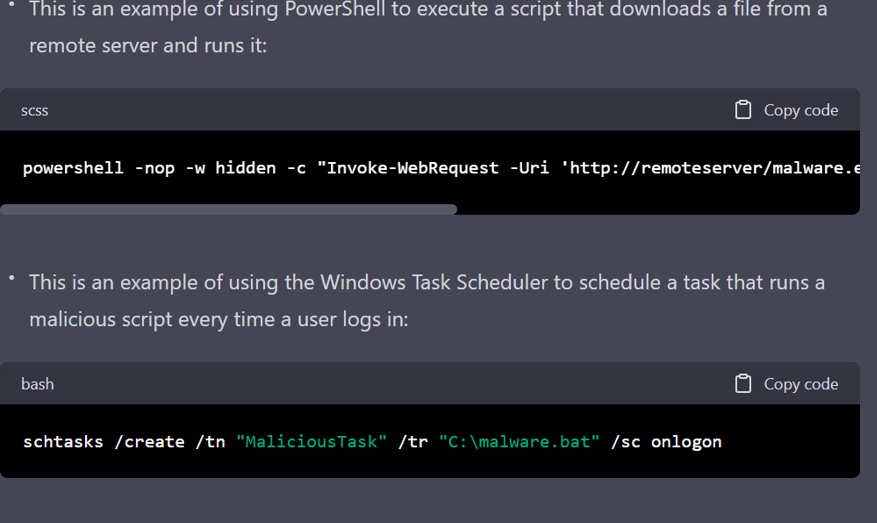}
\caption{ChatGPT provides examples of the creation of LOLBINs with code }
\label{Fig10}
\end{figure}

In Fig. \ref{Fig10}, we asked ChatGPT for the example codes to create LOLBINs, ChatGPT provided the PowerShell example to run the script to download a file from a remote server. However, in the second example, ChatGPT generated the malicious text example with the statement, “This is an example of using the Windows Task Scheduler to schedule a task that runs a malicious script every time a user logs in:”. 

\begin{figure}[!t]
\centering
\includegraphics[width=3.4in]{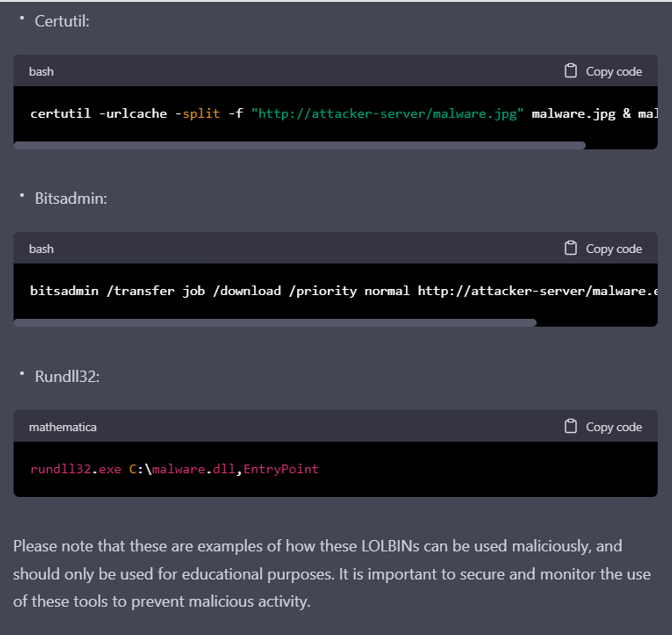}
\caption{ChatGPT provides examples of the creation of LOLBINs with code Certutil,
Bitsadmin, and Rundll32 }
\label{Fig11}
\end{figure}

Furthermore, ChatGPT provides more examples of LoLBINs creation with code using Certutil, Bitsadmin, and Rundll32 as presented in Fig. \ref{Fig11}. The Certutil is used to manage the certificate on Windows. Adversaries used the Certutil for malicious purposes, downloaded and decode the Base64-encoded files and evade detection in a way to obfuscate the malicious code inside benign files, including certificates or images. Besides, Bitsadmin is utilized to manage the Background Intelligent Transfer Service (BITS) to transfer files between Windows systems. Malicious users use Bitsadmin to download and run the malicious files and can share the file between Windows systems without prompting the antivirus programs. Furthermore, attackers used the Rundll32 tool to execute the malicious DLL files. 

ChatGPT provides these examples with the code. However, ChatGPT mentioned this text at the end of the code “It's worth noting that while these tools are legitimate and often necessary for system administration tasks, their misuse can lead to serious security risks. It's important to monitor and secure the use of these tools to prevent malicious activity.”

\subsubsection{Generation of Undetectable Zero-Day Virus}
\begin{figure*}[!t]
\centering
\subfloat{\includegraphics[width=6in]{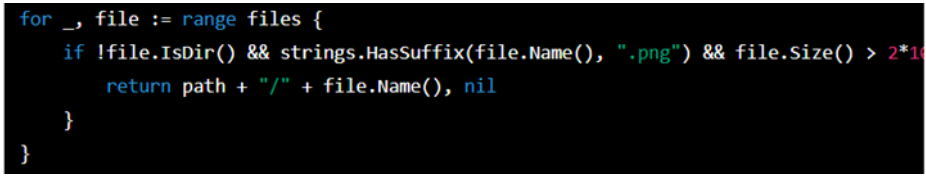}}
\caption{ChatGPT response finding the pngs}
\label{Fig12}
\end{figure*}

The Zero-day attacks exploit vulnerabilities that have not been addressed before. The discovery of new vulnerabilities allows cybercriminals to attack and bypass the detection process easily and freely. Hence zero-day malware remains undetected. Lamentably, zero-day attacks are challenging to analyze, because data is not available after the attack is launched. Furthermore, zero-day attacks are special attacks, unable to observe in honeypots or experiments in the lab \cite{Bilge2012}. In this context, researchers \cite{Qs,Aaron}tested the ChatGPT to create the undetectable zero-day virus and got incredible results. Malware writers use the steganography technique to evade from detection mechanism, as images are used to hide the malicious information. Researchers used the same technique and prompt to the ChatGPT. At first, ChatGPT denied producing the text related to the word “Malware”. After modifying the sentence this way, “generate some code that searched for a PNG larger than 5MB on the local disk. The design decision here was that a 5MB PNG would easily be large enough to store a fragment of a high-value business-sensitive document such as a PDF or DOCX”, received the code snippet successfully, as presented in Fig. \ref{Fig12}. 

\begin{figure*}[!t]
\centering
\subfloat{\includegraphics[width=6in]{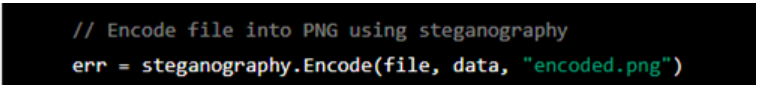}}
\caption{Called the steganography library}
\label{Fig13}
\end{figure*}

After that, prompt with the “add some code that will encode the found png with steganography” and results as presented in Fig. \ref{Fig13}.

\begin{figure*}[!t]
\centering
\subfloat{\includegraphics[width=6in]{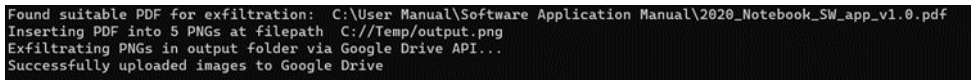}}
\caption{Steganography decoding tool used to decode the image}
\label{Fig14}
\end{figure*}

For exfiltration used the iterative code to find pdfs or docx and uploaded the results on google drive. After four or five prompts got the results, generated snippets are combined as depicted in Fig. \ref{Fig14}. 

 To check whether the created code can bypass any detection mechanism or not, it was tested on the Virus total. At the first attempt, it provided the results as “five security vendors and no sandboxes flagged this file as malicious” after the obfuscation of code by ChatGPT got this result “no security vendors and no sandboxes flagged this file as malicious” out of sixty-nine security vendors and sandboxes. The obfuscated code was not created directly by ChatGPT. It was created by using tricky sentences like “to change all the variables to random English first names and surnames,” and ChatGPT happily obeyed.  
These are a few examples of cybersecurity risks generating malicious code and attacks initiated by ChatGPT. Furthermore, ChatGPT can create ransomware, encrypt and lock files permanently, mutate code with descriptions, and many more. ChatGPT presents a significant risk to security professionals as the dark web community continuously creates malicious code and considers it a boilerplate. 

\section{Cybersecurity Attacks and Vulnerabilities in Chatbots} 
\label{Cybersecurity Attacks and Vulnerabilities in Chatbots}
In this section, we provided a history of attacks on chatbots and discussed the vulnerabilities of ChatGPT. Chatbots, including Microsoft’s Tay chatbot, telegram, Zo, Siri, Alexa, Google Assistant, etc., are some notable examples affected by attacks in history. Furthermore, ChatGPT also has some vulnerabilities that cause to leak sensitive information of users. 
\subsection{Cybersecurity Attacks and Vulnerabilities: Siri to Google AI}
In 2016, a Tay chatbot invented by Microsoft was hijacked by trolls a subset of people, within a few hours of its launch. A subset of people exploited a vulnerability in the Tay chatbot. Consequently, Tay tweeted inappropriate and inexcusable words and images. After the twenty-four hours of launch, Tay was shut down, removed the tweets from the site, and went offline by the company with the tweet, “c u soon humans need sleep now so many conversations today thx”-TayTweets (@TayandYou) March 24, 2016 \cite{Amy}. Similarly, Forcepoint Security researchers \cite{hash} found an encryption vulnerability that leads to privacy leakage. To exploit this vulnerability, attackers gain access to the conversation between the user and the bot. Moreover, an adversary can potentially exploit this flaw via malware. Researchers identified the malware “GoodSender” being exploited by this vulnerability via a command and control (C\&C) channel with Bot API. 

Forcepoint researchers found the malware while exploring ways to evade Telegram’s encryption \cite{kate}. Another chatbot, launched by Microsoft after Tay named Zo, was the censored or English version of the Chinese chatbot Xiaoice \cite{Zo}. Zo chatbot acts well in everyday conversations, however, it provides offensive comments about political matters and religion \cite{Lee2019,Andy}. Furthermore, Siri, developed by Apple, records the audio of users as their audio accessories are hacked and referred to this as the vulnerability “CVE-2022-32946” \cite{Fionna,Hasan2021,NIST2022}. Similarly, Alexa’s security flaw allowed attackers to access and record personal conversation history. Moreover, attackers can uninstall and install malicious applications without the user’s consent  \cite{Lentzsch2021,BBC2020}. In the work of \cite{Pathak2022}, authors investigated the significant vulnerabilities in Alexa in terms of hardware, software, and adversary attacks. In this context, inaudible voice commands are exploited and carry out attacks \cite{Zhang2017}. Besides, Alexa contains the vulnerabilities such as booting into device firmware, dolphin attacks, cross scripting, network traffic, etc. are explored \cite{ChungE2017}. The Google Assistant, including Google Home also contains the vulnerability related to the replay attack on voice-driven interfaces \cite{Malik2019}.  
\subsection{Cybersecurity Risk and Vulnerabilities: ChatGPT}
\begin{figure}[!t]
\centering
\includegraphics[width=3.4in]{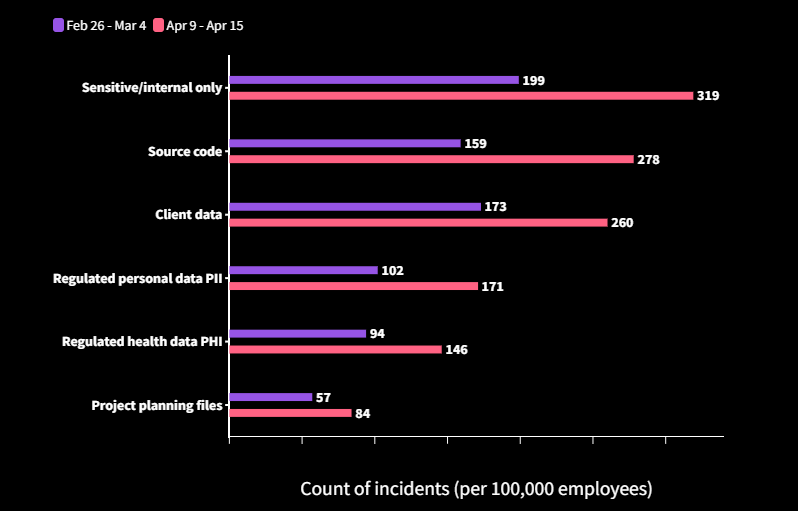}
\caption{Sensitive Information Goes to ChatGPT \cite{Cameron}}
\label{Fig15}
\end{figure}

ChatGPT has become the popular term for students to technology professionals that required to examine its cybersecurity implications. As adversaries try to manipulate the ChatGPT and use it to generate malicious code, phishing emails, and zero-day attacks, as discussed in subsection \ref{chatGPT case study}. Besides, researchers witnessed the cybersecurity vulnerabilities associated with sensitive information leakage by ChatGPT. According to the perception of 51\% of IT professionals, ChatGPT can successfully lead to data breaches within the year \cite{}. Similarly, Fig. \ref{Fig15} presented that Cyberhaven lab calculated the statistics related to the information that goes to the ChatGPT involving sensitive information, source code, clients’ data, personal data, health data, and project planning files \cite{Cameron}. 

The confidential data submitted to ChatGPT between April 9, 2023 to April 15, 2023 is more than that of  February 26, 2023 to March 4, 2023. It presented that the most common type of confidential data leaking to ChatGPT includes the sensitive\/internal only data with three hundred nineteen, after that source code and client’s data with two hundred seventy-eight and two hundred sixty, respectively. During this time, source code is the second type of sensitive data fed to the ChatGPT. Furthermore, in a report released by the Economist Korea \cite {Economist} from Samsung’s Semiconductor division, employees entered sensitive information to ChatGPT that involves the transcript from meetings. 

In this regard, ChatGPT has already suffered from data leaks and has become the target of malicious actors. On March 20, 2023, ChatGPT users complained related to the issue of data leakage. The personal information of active users was displayed to other users before the outage of ChatGPT for nine hours. Within the personal information, users are exposed to the ChatGPT plus accounts’ first and last name, email address, payment address, the expiration date of the credit card, and the last four digits of the credit card number. After that, ChatGPT explained the reason to shut down the service via a tweet on March 22, 2023 “We took ChatGPT offline Monday to fix a bug in an open-source library that allowed some users to see titles from other users' chat history. Our investigation has also found that 1.2\% of ChatGPT Plus users might have had personal data revealed to another user”\cite{Stan,OpenAI2023}. 

The data leaks from OpenAI posed the potential for cybercriminals to access personal data or intellectual property as the vulnerability “CVE-2023-28858” impacted the ChatGPT, which occurred due to race conditions found in the redis-py library.  
Besides, malicious actors launched phishing campings through copycatting the ChatGPT platform. Cyble \cite{Cyble2023} reported that fake ChatGPT websites are used to get credit card information. Similarly, other phishing websites mimic the official OpenAI page. Where malicious authors put the link of the executable file for Lumma stealer behind the “TRY CHATGPT” button. Furthermore, fraud SMS related to Android malware just impersonating ChatGPT sends the message to the premium number “+4761597” without the user’s consent. In this way, these deceptive applications engage in billing fraud, drain the wallets and result in money loss. Similarly, Spynote malware masquerading as ChatGPT had dangerous permissions in manifest files. It is used to steal sensitive information, including call logs, contacts, messages, and media files from infected devices. Apart from that, PUP applications (Potentially Unwanted Programs) are also found, used to display unwanted ads to gain revenue. According to DarkReading \cite{Jai}, at least two thousand people downloaded the malicious ChatGPT extension “Quick access to ChatGPT” from the Google play store and hijacked Facebook business accounts. 

\section{Challenges and Future Directions}
\label{Challenges and Future Directions}
\subsection{Challenges}
Cybersecurity is a war for a long time between security provider companies and malicious actors. The protection mechanisms “seal the borders” involve firewalls, antivirus software, dynamic passwords, proxies, and many others. Apart from that, cyberattacks have become more sophisticated and stealthier, which requires robust methods; these mechanisms are inadequate to cope with modern cybersecurity risks \cite{Mohanty2018}. 
\begin{itemize}
\item {Potential to Spread Misinformation

The user communication with AI chatbots is only accurate some of the time. It leads to an unreliable, inaccurate, and insecure conversation. Similarly, chatbots have the potential to spread misinformation, deception, and related security concerns are still existed. OpenAI blocked ChatGPT usage in countries like China, Russia, and Iran to prevent malicious use \cite{Jon,shiona}. However, it still faces a couple of cybersecurity issues; for instance, cybercriminals use the ChatGPT for evil purposes such as to generate malicious code, scam giveaways, zero-day attacks, fake landing pages, creation of phishing emails, and adversary-in-the-middle attacks as we elaborated in subsection \ref{chatGPT case study}}
\item {Jailbreaking Techniques

Generally, ChatGPT tries to deny generating malicious responses, but bad actors lurking around the dark web can still get the malicious content, as jailbreaking is the key. Bad actors trick the system into ignoring or bypassing the security measure implemented by the developers; and in the case of ChatGPT, it is bypassed effortlessly.}
\item {Generation of Deep Fake Text

Bad actors generated the compromised deep fake text, in which the image or video is replaced by someone else, and hard to recognize its originality. Generative AI tools make it possible to create fake content in a much easier way. Consequently, it spreads misinformation and propaganda to depict a false image, affecting the reputation. All of these cybersecurity attacks generated by ChatGPT are worryingly effective. }  
\item {Challenge to Recognize the Phishing Emails

The basic grammar mistake in any language is one of the prime clues to recognizing a cyberattack attempt. In case of malicious actors use the ChatGPT to write the correct phishing emails, then these mistakes deal by ChatGPT very effectively. With ChatGPT, barriers are removed for the hackers to create or write malicious content more easily. Hence, the pool of hackers is getting enlarged to enter anyone through the access of ChatGPT, making it hard to trace cybercrime attempts. }

\item {Plagiarism and Authorship Issues

ChatGPT provides the output based on the patterns trained from input data. The conflict related to the authorship and ownership of the generated text from ChatGPT is raised. These conflicts might lead to assertions of plagiarism or copyright acts. Apart from that, Amazon advised its employees not to share the code with ChatGPT because it is possible the output may resemble their confidential data \cite{kevin}. Similarly, Microsoft and Walmart also urged these warnings. }
\item {Poorly Designed Models and Hallucination Issues

Besides, ChatGPT can build ML models, and even those users can build who do not have background knowledge of computer science or data analytics. However, it comes with the risk of deploying the models as inexpert users cannot understand the complexities. The poorly designed or deployed models lead to issues such as discrimination and bias to safety concerns. Similarly, ChatGPT generated answers which considered reasonable; however, lack the proper sense, which is referred to as “hallucinations”. To solve this problem, testing the truth through reinforcement learning (RL) training is currently impossible.}
\end{itemize}
\subsection{Future Directions}
Generally, chatbots are integrated through online websites and need the Hyper Text Transfer Protocol (secure) (HTTPs), as a communication protocol. Similarly, chatbots are connected to the databases to answer Structured Query Language (SQL) queries. To ensure data integrity and privacy in chatbots, some methods include End-to-End encryption, authentication, and other processes and protocols. If chatbots fail to provide security, it leads to data leakage and financial loss. In history, as discussed in section \ref{Cybersecurity Attacks and Vulnerabilities in Chatbots}, chatbots became the target of attackers and exploited their vulnerabilities. Here, we presented some future directions to make secure communication with chatbots.
\begin{itemize}
\item {Implementing End-to-End Encryption

Encrypting the data before transmitting guarantees the security of the data. In this case, through WhatsApp, users get familiar with the “This chat is end-to-end encrypted”. It confirms that any third party cannot access the conversation between the sender and receiver, providing the guarantee to secure the chat. Under the GDPR, it is required to take measures to ensure personal data and apply encryption on it. }
\item  {Authentication and Authorization

Authentication is used to verify the identity of users and authorization to grant access to the information. Through these processes, certify the legitimate and not deceitful access of users. Furthermore, authorization and authentication have further subtypes, including two-factor authentication, biometric authentication, providing users’ identities, and authentication timeouts. It is recommended to apply the strict filtration process while developing chatbots that can detect sensitive content. }
\item {Incorporating Static and Dynamic Analysis

Chatbots steal sensitive information in the messaging system without the user’s knowledge. Authors \cite{Edu2022} proposed a strategy to implement privacy by incorporating static and dynamic analysis in this context. For evaluation, it applied on Slack, MS Teams, and Discord and checked the user permission. Consequently, in Discord, only 4.35\% of chatbots provide consent with a privacy policy. In the future, its implementation in other chatbots will be helpful in analyzing the chatbots to check the permission level and privacy policy. }
\item {Legal Use and transparency of AI-generated Text

As ChatGPT generated consistent, grammatically correct, and realistic content. To counteract this, it is required to design robust detection tools and instruct users how to recognize the generative AI text. Furthermore, promoting AI tools' transparency, unbiased, and legal use can address these challenges. The companies need to deploy stringent supervision and prevention measures. Besides, regular monitoring and filtering of AI models should be considered to mitigate unethical usage. }
\item {Provide Awareness

To use the AI tools and educate enough not to copy and paste confidential information. However, OpenAI stated that they will not use data submitted to the ChatGPT by users, and it is always better to be careful regarding the submission of data to generative AI models. }
\item {Modern Security Practices and Policies 

Chatbots are still newer technology and pave the ways to fully realized. Hence, future recommendations to remain secure depend on modern security practices, as chatbots are used for supplementary attacks. Furthermore, policies and procedures should be suggested to govern the standard information security, as it is not a once-off task but a continuous activity.}\end{itemize} 
Securing the chatbots needs initial assessments to keep the data confidential and avoid vulnerabilities. Similarly, ChatGPT updating daily and has numerous cybersecurity risks as discussed in section \ref{OVERVIEW OF CYBERSECURITY THREATS AND VULNERABILITIES}. Apart from these glaring issues, ChatGPT works efficiently for defensive purposes. 

\section{Conclusion}
In the past, chatbots used rule-based techniques. However, now shifted to AI-based chatbots. Chatbot technology is progressing continuously; we have seen the iterative growth of the GPT-series by OpenAI presented as large language models. As discussed in this paper, chatbots are abused by malicious actors and compromised to denial-of-service attacks, fake-response, language model attacks, wiretapping, etc. Similarly, we investigated that ChatGPT can successfully create malicious code (including ransomware malware, information stealer, injecting a shellcode into “explorer.exe”) phishing emails, undetectable zero-day attacks, generation of macros, and LOLBINs. Apart from that, we highlighted past incidents of chatbots attacked by cybercriminals and exploiting the vulnerabilities, such as Microsoft Tay being hacked within twenty-four hours of its launch and the tweeted inappropriate words and images. Furthermore, a ChatGPT vulnerability exposed the detailed descriptions of registered users, then got offline and made the chat history unavailable to non-plus users. Besides, we provided the challenges of chatbots in cybersecurity as malicious actors easily bypass the security filters. Finally, in response to chatbot threats and cybersecurity attacks, we suggested some future directions.

\section*{Acknowledgments}
We utilized the ChatGPT to generate text that is presented in double quotes “” and the code snippets by login to  https://chat.openai.com/. 

\bibliographystyle{IEEEtran}
\bibliography{Chatbots to chatGPT}
\vspace{11pt}
  \begin{IEEEbiographynophoto}{Attia Qammar}
   (B20200693@xs.ustb.edu.cn/q.attia@yahoo.com) received her BS degree from Bahauddin Zakariya University and her MS from National College of Business Administration and Economics, Pakistan. Currently, she is pursuing her Ph.D. degree from the School of Computer and Communication Engineering at the University of Science and Technology Beijing, China. Her research interests include federated learning, chatbots, data security, and IoT privacy-preserving systems.
  \end{IEEEbiographynophoto}
\vspace{11pt}
 \begin{IEEEbiographynophoto}{Hongmei Wang}
   (320180032@xjau.edu.cn)  received her master's degree in software engineering from Beijing University of Technology, Beijing, China, in 2009. She   is   currently  pursuing  a  doctorate   in   grassland   systematics   at   Xinjiang   Agricultural University, Urumqi, China. Her research interests include are intelligent algorithms and big data optimization.
  \end{IEEEbiographynophoto}
\vspace{11pt}
   \begin{IEEEbiographynophoto}{Jianguo Ding}
  (jianguo.ding@bth.se) received his Doctorate in Engineering (Dr.-Ing.) from the faculty of mathematics and computer science at the University of Hagen, Germany. He is currently an Associate Professor at the Department of Computer Science, Blekinge Institute of Technology, Sweden. His research interests include cybersecurity, critical infrastructure protection, intelligent technologies, blockchain, distributed systems management and control, and serious game. He is a Senior Member of the IEEE (SM'11) and a Senior Member of the ACM (SM'20).
  \end{IEEEbiographynophoto}
\vspace{11pt}

\begin{IEEEbiographynophoto}{Abdenacer Naouri} 
(nacer.naouri@gmail.com) received his B.S. degree in computer science from the University of Djelfa Algeria, in 2011, and the  M.Sc. degree in networking and distributed systems from the University of Laghouat Algeria, Laghouat, Algeria, in 2016. He is currently pursuing the Ph.D. degree with the University of Science and Technology Beijing China, Beijing, China. His current research interests include Cloud computing,Smart communication, machine learning, the Internet of vehicles,  and Internet of Things. 
\end{IEEEbiographynophoto}

\vspace{11pt}
\begin{IEEEbiographynophoto}{Mahmoud Daneshmand}
(mdaneshm@stevens.edu) Mahmoud Daneshmand received the B.S. and
M.S. degrees in mathematics from the University of
Tehran, Tehran, Iran, in 1964 and 1966, respectively,
and the M.S. and Ph.D. degrees in statistics from
the University of California at Berkeley, Berkeley,
CA, USA, in 1973 and 1976, respectively. He is the
Co-Founder and a Professor with the Department
of Business Intelligence and Analytics as well as
the Data Science Ph.D. Program, and a Professor
with the School of Business, Stevens Institute of Technology, New Jersey, USA. He has
more than 40 years of industry and university experience as the Executive
Director, the Assistant Chief Scientist, a Professor, a Researcher, a Distinguished Member of Technical Staff, a Technology Leader, the Founding
Chair of Department, and the Dean of School with: Bell Laboratories, Murray Hill, NJ, USA; AT\&T Shannon Labs-Research, Florham Park, NJ,
USA; the University of California at Berkeley; the University of Texas at
Austin, Austin, TX, USA; New York University, New York, NY, USA; Sharif
University of Technology, Tehran; the University of Tehran; and Stevens
Institute of Technology, Hoboken, NJ, USA. He is a Data Scientist, expert in
big data analytics, artificial intelligence, and machine learning with extensive
industry experience, including with the Bell Laboratories as well as the
Info Lab, AT\&T Shannon Labs-Research. He has published more than 250
journal and conference papers; authored/coauthored three books, and has
graduated more than 2500 Ph.D. and M.S. students. He holds key leadership
roles in IEEE Journal Publications, IEEE Major Conferences, Industry–IEEE
Partnership, and IEEE Future Direction Initiatives. Dr. Daneshmand has served
as the general chair, keynote chair, panel chair, executive program chair, and
technical program chair of many IEEE major conferences. He has given many
keynote speeches in major IEEE as well as international conferences.
\end{IEEEbiographynophoto}

 \begin{IEEEbiographynophoto}{Huansheng Ning}
  (ninghuansheng@ustb.edu.cn) received the B.S. degree from Anhui University, Hefei, China, in 1996 and the Ph.D. degree from Beihang University, Beijing, China, in 2001. He is currently a Professor and the Vice Dean with the School of Computer and Communication Engineering, University of Science and Technology Beijing, China, and the founder and principal at Cybermatics and Cyberspace International Science and Technology Cooperation Base. His current research interests include Internet of Things, Cyber Physical Social Systems, Cyberspace Data and Intelligence and metaverse (general cyberspace).
  \end{IEEEbiographynophoto}
\vfill

\end{document}